\documentclass[aps,prx,twocolumn,groupedaddress,amsmath,epsfig,amssymb,eqsecnum]{revtex4}
\usepackage[utf8]{inputenc}
\usepackage[english]{babel}

\usepackage{epsfig,amsmath,amssymb,bm,epsf,graphics,graphicx,psfrag,verbatim,subfigure,framed}
\usepackage{makecell}
\usepackage{bbm}
\usepackage{mathrsfs}
\usepackage{amsfonts}
\usepackage{color}
\usepackage{graphicx}
\newcommand{\xing}[1] {{\color{red} #1 }}

\newcommand{\xv}{\boldsymbol x}

\newcommand{\qv}{{\boldsymbol q}}
\newcommand{\pv}{{\boldsymbol p}}
\newcommand{\yv}{{\boldsymbol y}}

\newcommand{\Tr}{{\rm Tr}}

\newcommand{\Am}{{\boldsymbol{A}}}
\newcommand{\Om}{{\boldsymbol{O}}}
\newcommand{\Qm}{{\boldsymbol{Q}}}

\newcommand{\Lm}{{\boldsymbol L}}

\newcommand{\Bm}{{\boldsymbol B}}

\newcommand{\be}{\begin{equation}}
\newcommand{\ee}{\end{equation}}
\newcommand{\ba}{\begin{eqnarray}}
\newcommand{\ea}{\end{eqnarray}}
\begin{document}
\title{Covariant Formulation of {Non-linear} Langevin Theory \\
with Multiplicative Gaussian White Noises}
\author{Mingnan Ding$^{1}$}
\author{Zhanchun Tu$^{2}$}
\author{Xiangjun Xing$^{1,3,4}$}
\email{xxing@sjtu.edu.cn}
\address{$^1$Wilczek Quantum Center, School of Physics and Astronomy, Shanghai Jiao Tong University, Shanghai 200240, China \\
$^2$Department of Physics, Beijing Normal University, Beijing 100875, China\\
$^3$T.D. Lee Institute, Shanghai Jiao Tong University, Shanghai 200240, China\\
$^4$Shanghai Research Center for Quantum Sciences, Shanghai 201315, China}
\date{\today} 


\begin{abstract} 
{The multi-dimensional} non-linear Langevin equation with multiplicative Gaussian white noises in Ito's sense is made covariant with respect to non-linear transform of variables.  The formalism involves no metric or affine connection, works for systems with or without detailed balance, and is substantially simpler than previous theories.  Its relation with deterministic theory is clarified. The unitary limit and Hermitian limit of the theory are examined.  Some implications on the choices of stochastic calculus are also discussed.  


\end{abstract}
\maketitle 

\section{ Introduction}   

Non-linear Langevin theory with multiplicative noises~\cite{Gardiner-book,Jacobs-book,van-Kampen-stochastic,Hanggi-review-1982}  is widely used to describe dynamics out of equilibrium.  Yet  development of this theory turns out to be very challenging and full of controversies.  There have been extensive and long-lasting discussions on the choice of stochastic calculus~\cite{Germano2009,Kampen-1981,Klimontovich-1990,Kupferman-2004,Sokolov-2010,Mannella-2012}, relation between deterministic and stochastic description~\cite{Hanggi-1980,Hanggi-1981,van-Kampen-validity}, as well as the covariance of theory under non-linear transform of variables (NTV)~
\cite{Graham-covariance-Langevin,Graham-covariance-Langevin-Ito,Grabert-1979,Grabert-1980s}. Another related issue is discretization scheme for its path integral representation~\cite{Lau-Lubensky-2007,Langouche-funct-integ-book,Arnold-2000,Arnold-2000-2,Arnold-2000-2-comment,Barci-2015,Barci-2016,Cugliandolo-path-integral}.
 Up to now, non-linear Langevin theory with multiplicative noises is  deemed not yet properly understood.  While earlier works mostly focus on processes with detailed balance (DB), more recently there have been many efforts trying to develop non-linear Langevin dynamics lacking DB~\cite{PAo-2004,KAT-2005,Jarzynski-2006,Xing-Jianhua-2010,Qian-Hong-decomposition,Qian-Hong-2014,Wang-Jin-2015-review}. Nonetheless, various  conceptual issues frequently come back.

We think that covariance (with respect to NTV) and DB are two issues of great importance, not only to Langevin theory, but also to non-equilibrium statistical mechanics at large. The common theme of non-equilibrium statistical mechanics is dynamics of {slow variables}.  But nonlinear functions of slow variables are also slow variables. (Of course, we need to assume that these functions are not fast themselves, i.e., they are not ``crazy''.)  This implies that statistical mechanics must be covariant under NTV.  When solving high dimensional problems, NTV is extremely useful, and even indispensable.  Also approximations must be covariant, or they lead to inconsistency.  More importantly, DB must be covariant under NTV, or an equilibrium system would be transformed into a non-equilibrium one, which is total nonsense.  Finally, it is by now clear that entropy production can be quantified in terms of DB violation.  Hence  DB plays a role in non-equilibrium statistical physics much like speed of light in relativity.  Even though processes without DB do exist, those with DB are special and must be invariant under NTV.  
  

Covariance of nonlinear Langevin equation and of associated Fokker-Planck theory was first addressed by Stratonovich~\cite{Stratonovich-covariance-1968}, and studied in more detail by Graham (first in Stratonovich's sense~\cite{Graham-covariance-Langevin}, and later in Ito's sense~\cite{Graham-covariance-Langevin-Ito}).  Graham's theory involves metric and affine connections and is very complicated, which hampered its application.  Soon after Grabert, Graham, and Green~\cite{Grabert-1980s} (GGG) greatly simplified the covariant formulation of Fokker-Planck theory.  The same problem was also addressed by H\"anggi~\cite{Hanggi-review-1982}.  Later Ramshaw~\cite{augmented-Langevin} showed how to derive GGG's Fokker-Planck theory from Langevin theory,  but no conclusive result has been established about covariance of nonlinear Langevin theory.   For a more recent work on covariant formulation of Langevin theory involving metric and affine connections, see reference~\cite{Polettini-Langevin-2013}.  


In this work we discuss a covariant formulation of non-linear Langevin theory which involves no metric tensor or affine connection.  Rules of transformation for parameters and physical variables are clearly demonstrated.  The theory cannot be simplified further without loss of generality or physical meanings.  Fully consistent deterministic limit can be obtained as either the thermodynamic limit or the low temperature limit.  While our formalism is inspired by the previous works, it is substantially simpler and more general, and more clear in physical meanings.  It works for systems with both even and odd variables, with or without detailed balance.  We believe that it is of considerable value to the general theory of non-linear Langevin dynamics with multiplicative noises.  

The remains of this work is organized as follows.  In Sec.~\ref{sec:form} we derive the covariant form of Langevin theory and Fokker-Planck theory, and explicitly demonstrate their covariance under general nonlinear transformation of variables. We also discuss the deterministic limit, unitary limit, and Hermitian limit of our theory, which are again covariant.  In Sec.~\ref{sec:DB}, we discuss time-reversal symmetry of the theory, and formulate the conditions of detailed balance in a fully covariant fashion.  In Sec.~\ref{sec:conclusion} we conclude this work with some comments on the general issue of stochastic calculus. 


\section{ Covariant Form} 
\label{sec:form}

In the traditional {\em Langevin approach}~\cite{van-Kampen-stochastic}, one starts from deterministic equation for slow variables and adds noises to obtain a Langevin equation.
While this approach works perfectly for linear systems, it leads to many controversies in the nonlinear case~\cite{van-Kampen-stochastic,Hanggi-1980,Hanggi-1981,Grabert-1980s}.  Here we will start with Langevin theory as a purely phenomenological theory, and re-express it in terms of observable quantities, such that it becomes fully covariant.   After understanding the deterministic limit of our theory, we will see why the conventional approach is so difficult.   


 Let $\xv = (x_1, \ldots, x_n)$ be the slow variables, whose dynamic evolution obey a stochastic differential equation.  The most general non-linear Langevin equation is either of the following two equivalent forms:
\begin{subequations}
\label{NL-Langevin-1-1}
\ba
dx_i &=& F_i( \xv, t) dt + b_{i \alpha}(\xv)  d W_\alpha(t),
\label{NL-Langevin-1-1-2} \\ 
\dot x_i &=&  F_i(\xv,t) +  b_{i \alpha}(\xv) \xi_\alpha(t).  
\label{NL-Langevin-1-1-1}
\ea 
\end{subequations} 
where repeated indices are summed over.    In the above $W_\alpha(t), \alpha  = 1, \cdots, m$  are $m$ dimensional Wiener process, whereas $ \xi_\alpha(t) = d W_\alpha(t)/dt$ are  normalized Gaussian white noises with correlations given by
\ba
 \langle \xi_\alpha(t)\xi_\beta (t')\rangle
  &=&  \delta_{\alpha \beta} \delta (t - t').
  \label{xi-correlation}
\ea
 We expect $m \gg n$, since there are in general many more fast variables (which are modeled as white noises in this theory) than slow variables.   In this work, we shall assume that $F_i(\xv,t)$ and  $b_{i \alpha}(\xv)$ are both time-independent.   The products $b_{i \alpha}(\xv) dW_\alpha(t)$ and $ b_{i \alpha}(\xv) \xi_\alpha(t)$ in Eq.~(\ref{NL-Langevin-1-1}) are defined in Ito's sense.   Finally while Eq.~(\ref{NL-Langevin-1-1-2}) is the mathematically rigorous formulation of stochastic differential equation and universally used by mathematician, Eq.~(\ref{NL-Langevin-1-1-1}) is the traditional formulation by Langevin~\cite{Langevin-Lemons-review}, and is still preferred by many physicists.  
 We refer to the classical textbook by Gardiner \cite{Gardiner-book} for a detailed introduction of all relevant formalisms. 
 

The Fokker-Planck equation (FPE) associated with Langevin equation (\ref{NL-Langevin-1-1}) can be derived using standard method~\cite{Gardiner-book,Jacobs-book}:
\ba
 \partial_t p(\xv, t)  &=& 
- \partial_i \left( F_i (\xv, t)p(\xv, t)   \right) 
+ \partial_i \partial_j \left( B_{ij} (\xv)p(\xv, t)  \right)
\nonumber\\
&=& - \partial_i j_i(\xv,t),
\label{FP-1}
\ea
where the matrix $\Bm = {(B_{ij})}$ is given by
\be
B_{ij} (\xv) \equiv
 \frac{1}{2}\, b_{i \alpha} (\xv) b_{j\alpha} (\xv)
 = B_{ji}(\xv). 
 \label{B-b-def} 
\ee    
Since $ b_{i \alpha} (\xv)$ are real, the matrix $\Bm$ is non-negative.  It however may have zero eigenvalues.  ${j_i}(\xv,t)$ is the  {\em probability current} defined as
\be
j_i \equiv   F_i p - \partial_j \left( B_{ij} p\right).
\label{prob-current-def}
\ee 
Equation (\ref{FP-1}) has the form of continuity equation: $\partial_t p + \partial_i j_i=  0$, which describes conservation of probability.  

We shall assume that the Langevin dynamics has a unique steady state 
\be
 p^S(\xv) = e^{-U(\xv)}, 
 \label{P-S-U-def}
\ee  
where $U(\xv)$ is called the {\em generalized potential}.  Substituting Eq.~(\ref{P-S-U-def}) in Eq.~(\ref{prob-current-def}), the steady state probability current can be re-expressed as
\ba
j^S_i &=& ( F_i - ( \partial_jB_{ij}) + B_{ij}( \partial_j U))e^{-U}. 
 \label{F^S-1}
\ea
 Using this result to express $F_i$ in terms of $j^S_i$, we can rewrite Eq.~(\ref{FP-1})  into \xing{}
\be
\partial_t p =  - \partial_i (j^S_i e^{U} p) 
+ \partial_i B_{ij} (\partial_j  + (\partial_j U ) )p. 
\label{FP-2-1}
\ee
 Note that each of the two terms in  RHS vanishes separately in the steady state.  
 
Since the steady state current is conserved, $  \partial_i j_i^S = 0$, it is always possible to write it in terms of an antisymmetric matrix function $\Qm$ with components $Q_{ij} = - Q_{ji}$, such that
  \be
 j^S_i = \partial_j (Q_{ij} e^{-U}).\xing{} 
 \label{F^S-2}
 \ee
   Such a parameterization was first used by Graham~\cite{Graham-covariance-Langevin}, and later by Eyink et. al.~\cite{Eyink-1996} and also by J.H.Xing~\cite{Xing-Jianhua-2010}.  Note that $\Qm$ is generally not unique.  Using Eq.~(\ref{F^S-2}) in Eq.~(\ref{F^S-1}), we can express $F_i$ as\xing{}
   \be
F_i = - L_{ij} \partial_j U + \partial_j L_{ij}.    
\label{F-L-U-def}
   \ee   
where $\Lm = (L_{ij}) $ is the matrix of {\em kinetic coefficients} with
\be
L_{ij} = B_{ij} + Q_{ij}, \quad
B_{ij} = B_{ji}, \quad Q_{ij} = - Q_{ji},
\label{L-B-Q-decomp}
\ee
Substituting Eq.~(\ref{F-L-U-def}) back into Eqs.~(\ref{NL-Langevin-1-1}),  we obtain the following standard forms of non-linear Langevin equation:
\begin{subequations}
\ba
d x_i &=& 
- L_{ij}(\xv) \partial_j U (\xv) dt  + \partial_j L_{ij} (\xv) dt
+  b_{i \alpha}(\xv)  d W_\alpha(t),
\nonumber\\
\\
\dot x_i &=& 
- L_{ij}(\xv) \partial_j U (\xv) + \partial_j L_{ij} (\xv)
+  b_{i \alpha}(\xv)  \xi_\alpha(t),
\nonumber\\
\ea 
 \label{NL-Langevin-standard}
\end{subequations}
Substituting Eq.~(\ref{F-L-U-def}) back into Eqs.~(\ref{FP-1}), and (\ref{prob-current-def}), we obtain the following standard form of FPE:
\begin{subequations}
\be
\partial_t p = {\mathcal L}_{\rm FP} p
= - \partial_i j_i , 
\label{FP-3}
 \ee 
where the Fokker-Planck operator $ {\mathcal L}_{\rm FP} $ and the probability current $\textit{\textbf {j}}$ are given respectively by
\ba
{\mathcal L}_{\rm FP}
&\equiv& \partial_i L_{ij} (\partial_j  + (\partial_j U ) ),
\label{L_FP-def} 
\vspace{3mm}\\
j_i &=&  - L_{ij}(\partial_j + (\partial_j U)) p + \partial_j(Q_{ij} p ) .
\quad\quad
\label{prob-current-2} 
\ea
\end{subequations}
Note that Eq.~(\ref{FP-3})  has a formal solution $p(t)  = e^{ t {\mathcal L}_{\rm FP}} p(0)$.  Since ${\mathcal L}_{FP} e^{-U} = 0$, $p^S(\xv) = e^{- U(\xv)}$ is indeed the steady state solution.  


As one can see  from Eq.~(\ref{L_FP-def}), the Fokker-Planck operator depends on  $b_{i\alpha}(\xv)$ only through the combination $B_{ij}(\xv) = b_{i \alpha} (\xv) b_{j\alpha} (\xv)/2$, as defined in Eq.~(\ref{B-b-def}).  Consequently we can apply $\xv$-dependent right rotation on $b_{i\alpha}(\xv)$ without changing the dynamics  of $\xv$.  This freedom was first pointed out by Graham~\cite{Graham-covariance-Langevin-Ito}.


\subsection{\bf Proof of Covariance} 
First we define the notions of covariance and contravariance.  Suppose $\xv = (x_1, x_2, \ldots, x_n)$ are the slow variables, and we have a Langevin dynamics as formulated above.  Consider a time-independent NTV $\xv \rightarrow \yv = (y_1,\cdots, y_n)$ and let ${ J} \equiv \det ( \partial y_a/\partial x_i )$ be the Jacobian.   As we discussed in the introduction,  $\yv = (y_1,\cdots, y_n)$ are also slow variables, and hence the Langevin theory can also be formulated in terms of $\yv$, and two formulations must be mathematically equivalent.  As in general relativity theory, vectors and tensors appearing in two formulations are related to each other via linear transformations, whose coefficients are generically nonlinear functions of slow variables.  Unlike in general relativity theory, however, metric tensor plays no role in our Langevin theory, because there is no notion of distance in the manifold of slow variables.

Suppose a vector $z_i( \xv)$ and a tensor $A_{ij}(\xv)$ are transformed to  $z_a'(\yv)$ and $A'_{ab}(\yv)$ in the new coordinate system.  We call them {\em covariant} if they transform  as
\begin{subequations}
\ba
z'_a (\yv)& = &  \frac{\partial x_i}{\partial y_a} z_{i},
\\
A'_{a b } (\yv)& = & \frac{\partial x_i}{\partial y_a} 
A_{ij} (\xv)\frac{\partial x_j}{\partial y_b}.
\ea
\end{subequations}
 We call them {\em contra-variant} if they transform as  
 \begin{subequations}
\ba
z'_a (\yv) & = & \frac{\partial y_a}{\partial x_i} z_{i},
\\
A'_{a b } (\yv)& = & \frac{\partial y_a}{\partial x_i} 
A_{ij} (\xv)\frac{\partial y_b}{\partial x_j}.
\ea
\end{subequations}
Usually the indices of contra-variant vectors and tensors are displayed as superscripts instead of subscripts, whereas those of covariant vectors and tensors are displayed as subscripts.  Here to unclutter the notations, we display all indices as subscripts, and indicate explicitly how they transform under NTV.

We call a function $f(\xv)$ a {\em scalar} if it transforms as 
\be
f(\xv) \rightarrow f'(\yv) = f(\xv). 
\ee
We call a function $\phi(\xv)$ a {\em density} if it transforms as 
\be
\phi(\xv) \rightarrow \phi'(\yv) = { J}^{-1} \phi(\xv), 
\ee
which also implies $\phi'(\yv) d^n \yv = \phi(\xv) d^n \xv$.  

Below we will demonstrate that the nonlinear Langevin theory and Fokker-Planck theory, Eqs.~(\ref{NL-Langevin-standard}) and (\ref{FP-3}) respectively, keep their form under nonlinear transform of variables, but with all vectors and tensors replaced by their appropriately transformed versions (see Eqs.~(\ref{transform-U-L-b}) below).  This is what we mean by the {\em covariance of the nonlinear Langevin theory and Fokker-Planck theory}.  



While the old theory is characterized by $U(\xv)$, $L_{ij}(\xv)$, $b_{i\alpha}(\xv)$ as well as a probability {density} $p(\xv)$, the new theory is characterized by $U'(\yv), L'_{ab}(\yv), b'_{a\alpha}(\yv)$ as well as the probability {density} $p'(\yv)$.   We shall directly write down the rules of transform between two theories:
\begin{subequations}
\label{transform-U-L-b}
\ba
p'(\yv) &=& J^{-1}\, p(\xv). 
\label{covariant-rules-1}\\
U'(\yv) &=& U(\xv) + \log J, 
\label{covariant-rules-U}\\
b'_{a \alpha}(\yv) &=& \frac{\partial y_a}{\partial x_i}
b_{i \alpha}(\xv),
\label{covariant-rules-b}\\
L'_{ab}(\yv)  &= & \frac{\partial y_a}{\partial x_i} L_{ij} (\xv)
\frac{\partial y_b}{\partial x_j}.
\label{covariant-rules-3}
\ea 
Hence pdf $p(\xv)$ transform as a density,  whereas $b_{i \alpha}$ and $L_{ij}$ transform  as respectively contra-variant vector and tensor of rank two.  Note that the transform of generalized potential $U$ is such that the steady state $e^{-U(\xv)}$ transforms a density $e^{-U(\xv)} d \xv = e^{-U'(\yv)} d \yv$.  Note also that Eq.~(\ref{covariant-rules-3}) implies that $\Bm$ and $\Qm$ transform as
\ba
B'_{ab}(\yv)  &= & \frac{\partial y_a}{\partial x_i} B_{ij} (\xv)
\frac{\partial y_b}{\partial x_j}, \\
Q'_{ab}(\yv)  &= & \frac{\partial y_a}{\partial x_i} Q_{ij} (\xv)
\frac{\partial y_b}{\partial x_j}.
\label{Q'-Q}
\ea  
Hence $\Bm$ and $\Qm$ do not mix under NTV.   The Fokker-Planck operator and probability current in the transformed theory are obtained analogous to Eqs.~(\ref{L_FP-def}) and (\ref{prob-current-2})):
\ba
 {\mathcal L}'_{ \rm FP} &=&  {\partial'_a}
L'_{ab} \left( {\partial'_b} 
+ ( {\partial'_b} U')  \right) ,
\label{Fokker-Planck-2-c} \\
j'_a &=&  - L'_{ab}(\partial'_b + (\partial'_b U')) p'
 + \partial'_b(Q'_{ab} p' ), \quad 
 \label{J'-L'-P'}
\ea 
\end{subequations}
where we have used the shorthand $\partial'_a = \partial /\partial y_a$. 

In Appendix~\ref{sec:app_A}, we prove that under the above transform, Eqs.~(\ref{NL-Langevin-standard}) are transformed into 
\begin{subequations}
\label{dy-Ito-1-2}
\ba
 d y_a &=&   -  L'_{ab} \partial'_b U'  dt 
+ \partial'_b L'_{ab}  dt 
+ b'_{a \alpha}  dW_\alpha,\\
\dot y_a &=&   -  L'_{ab} \partial'_b U' 
+ \partial'_b L'_{ab} 
+ b'_{a \alpha}  \xi_\alpha .
\ea
\end{subequations}
 Ito's formula plays a crucial role in  proof of Eq.~(\ref{dy-Ito-1-2}). Moreover, we also prove that the transformed FP operator $ {\mathcal L}'_{ \rm FP}$ (Eq.~(\ref{Fokker-Planck-2-c})) and probability current $ j'_a$ (Eq.~(\ref{J'-L'-P'})) are related to those in the old theory $ {\mathcal L}_{ \rm FP}, j_i$ (which are respectively defined in Eqs.~(\ref{L_FP-def}) and (\ref{prob-current-2})) via
\begin{subequations}
\label{transformed-FP}
\ba
{\mathcal L}_{\rm FP}' &=& 
J^{-1} {\mathcal L}_{\rm FP} J,
\label{covariant-rules-5}\\
j'_a  &=& J^{-1} (\partial y_a/\partial x_i) j_i, 
\label{covariant-rules-4}\\
\partial'_a j'_a &=& J^{-1} \partial_i j_i .
\label{covariant-rules-4-1}
\ea
It then follows that 
\ba
{\partial_t p'} &=& J^{-1}\,{\partial_t }  p
= J^{-1} {\mathcal L}_{ \rm FP} p
\nonumber\\
&=& J^{-1} {\mathcal L}_{ \rm FP} JJ^{-1} p
=  {\mathcal L}'_{ \rm FP} p'
\nonumber\\
&=& - \partial'_a j'_a,
\ea 
where we have used Eqs.~(\ref{covariant-rules-1}), (\ref{FP-3}), and (\ref{covariant-rules-5}).  Hence we obtain the transformed FPE:
\ba
{\partial_t p'} =   {\mathcal L}'_{ \rm FP} p' =  - \partial'_a j'_a. 
\label{Fokker-Planck-2-b} 
\ea
\end{subequations} 
The covariance of our theory is evident by comparing Eqs.~(\ref{dy-Ito-1-2}) and (\ref{Fokker-Planck-2-b}) with  Eqs.~(\ref{NL-Langevin-standard}) and  (\ref{FP-3}).  We also note that Eq.~(\ref{covariant-rules-4-1}) implies that steady solutions keep their identity during NTV.  

Comparing with the earlier works~\cite{Graham-covariance-Langevin,Graham-covariance-Langevin-Ito,Grabert-1979,Grabert-1980s,Polettini-Langevin-2013}, our formalism does not involve affine connection or metric tensor, and establishes the covariance of Langevin theory and Fokker Planck theory simultaneously. {(The essence of metric is to define the notion of distance in the space of slow variables, which is not needed for study of statistical mechanics. Indeed, there were discussions on the {\em natural choice} of metric in earlier works~\cite{Graham-covariance-Langevin,Graham-covariance-Langevin-Ito,Grabert-1979,Grabert-1980s}, which did not lead to a clear cut conclusion.)} This makes our formalism much simpler and more useful.  Note also that our formalism is applicable regardless of DB.  


\subsection{Spurious Drift}

 The term $\partial_j L_{ij} (\xv)$ in the RHS of Eq.~(\ref{NL-Langevin-standard}) may appear unpleasant because it spoils the linear relation between the deterministic force and the {\em thermodynamic forces} $\partial_i U$.  Furthermore, because $\partial_j L_{ij} (\xv)$ generally contains a constant part, it also makes the most probable value of $\xv$ different from the minimum of $U(\xv)$.  We will call this term the {\em spurious drift}.   A part of this term $\partial_j B_{ij}$  has received lots of attention~\cite{Hanggi-1980,Lau-Lubensky-2007,Itami-Sasa-Langevin,van-Kampen-stochastic}.  The other part $\partial_j Q_{ij}$ though less well-known, is also important to guarantee the covariance of the theory.  

One might wish to remove the term $\partial_j L_{ij} (\xv)$ by choosing a different stochastic calculus, i.e., different interpretation of the dot product $b_{i \alpha}(\xv) \xi_\alpha(t)$.  As pointed out by H\"anggi~\cite{Hanggi-1980} for one-dimensional case long ago, in general neither Stratonovich nor kinetic can achieve the purpose.  For high dimension cases, such a stochastic calculus does not exist, because the antisymmetric matrix $\Qm (\xv)$ appears in $\partial_j L_{ij} (\xv)$ but not in the noise terms.  One might also attempt to absorb the term $\partial_j L_{ij} (\xv)$ by redefining the function $U(\xv)$, but then $U(\xv)$ would lose its contact with the stationary state, i.e., it is no longer equal to $- \log p^S(\xv)$.  Finally one might wish to make NTV such that the new symmetric matrix $B'_{ab}$ becomes independent of $\yv$.  This is generically impossible~\cite{Graham-covariance-Langevin} for the same reason that Riemann curvature cannot transformed away.  Furthermore, even if $B'_{ab}$ becomes constant, $Q'_{ab}$ are generically not. Hence Eq.~(\ref{NL-Langevin-standard}) cannot be simplified further without loss of generality or physical significance.  

\subsection{ Deterministic Limit} 
\label{sec:determinstic}
 When solving nonlinear Langevin dynamics, one may wish to take the {\em drift approximation} by throwing out the noise term in Eq.~(\ref{NL-Langevin-standard}), and obtaining a deterministic equation.  Such an approximation generically destroys the covariance, and therefore leads to inconsistency. {(One can straightforwardly repeat our proof of covariance in SI and demonstrate this point.)}  However if we delete both the spurious drift and the noise term in Eq.~(\ref{NL-Langevin-standard}), we obtain a deterministic equation:
\begin{subequations}
\ba
\dot x_i &=&  - L_{ij}(\xv) \partial_j U (\xv),
\label{determ-limit-1}
 \ea
which describes the irreversible dynamics of slow variables.  More precisely $\xv$ relaxes  towards the minimum of $U(\xv)$.  The relaxation may be oscillatory if $\Qm \neq 0$.  This equation is  covariant not under the transform as specified by Eqs.~(\ref{transform-U-L-b}), but under the following {\em revised rules of transform} for $U$:
\ba
U'(\yv) &=& U(\xv)  \label{rule-determ-limit-1}.
\ea
\label{determ-limit-all}
\end{subequations} 
The transform of $L_{ij}$ remains the same as in Eq.~(\ref{covariant-rules-3}).  
Proof of covariance is elementary.   Anticipating Eqs.~(\ref{reversibility-0}), we easily see that the deterministic theory (\ref{determ-limit-all}) satisfies detailed balance if and only if the stochastic theory does so.  


But how to reconcile the different transformation rules for $U(\xv)$, Eq.~(\ref{covariant-rules-U}) in the Langevin theory and Eq.~(\ref{rule-determ-limit-1}) in the deterministic theory?  In generic cases, this difference indicates there is no consistent way of taking deterministic limit of a stochastic theory.  However, there are two { important cases} where the inconsistency goes away.  The first case corresponds to the {\em thermodynamic limit}, where both $\xv$ and $U(\xv)$ are extensive, whereas spurious drift $\partial_j L_{ij} $, the Jacobian $J$ and the noises are all sub-extensive.   Throw out these sub-extensive terms, Eq.~(\ref{NL-Langevin-standard}) reduces to Eq.~(\ref{determ-limit-1}), and Eq.~(\ref{covariant-rules-U}) reduces to Eq.~(\ref{rule-determ-limit-1}), hence we obtain the deterministic theory Eqs.~(\ref{determ-limit-all}). 
 The second case corresponds to the {\em low temperature limit}, where $U = F/T$, with $F$ the free energy, and $T \rightarrow 0$.  
In this case, we have $\xv$ independent of $T$, whereas $L_{ij}$ linear in $T$.  The noises amplitudes $b_{i\alpha}$ then scale as $\sqrt{T}$ according to Eq.~(\ref{B-b-def}).  In this low temperature limit, again Eq.~(\ref{NL-Langevin-standard}) reduces to Eq.~(\ref{determ-limit-1}), and Eq.~(\ref{covariant-rules-U}) reduces to Eq.~(\ref{rule-determ-limit-1}), and hence we obtain the deterministic theory Eqs.~(\ref{determ-limit-all}).  

This discussion naturally leads us to the converse question: If we are given a deterministic dynamics $\dot x_i = F_i(\xv)$ together with a generalized potential $U(\xv)$, can we construct a covariant stochastic theory such that $e^{-U(\xv)}$ is the steady state?  This has been one of the perplexing questions in the {\em traditional Langevin approach}~\cite{van-Kampen-stochastic}.  We are now imposing it in a more general setting that is independent of detailed balance (DB).  The answer is {\em yes}, if we are willing to accept different transformation rules for $U$ in the deterministic and stochastic theories.  The matrix of kinetic coefficients can be found by solving the equations $L_{ij} \partial_i U= - F_i$.  Inserting them back to Eq.~(\ref{NL-Langevin-standard}) we find the desired covariant non-linear Langevin theory.  Without first understanding of the covariant theory Eq.~(\ref{NL-Langevin-standard}), however, it is very difficult to take such a leap from deterministic to stochastic!  



\subsection{\bf Unitary Limit and Hermitian Limit}  
\label{sec:unitary-Hermitian}
Following Qian~\cite{Qian-Hong-decomposition}, we define an inner product of two functions 
\be
(\phi, \psi )_{ U} \equiv  \int d^n\xv\,
e^{ U(\xv)}\, \phi(\xv) \psi(\xv)
= (\psi, \phi)_{ U},
\label{inner-prod-def}
\ee
where $ U(\xv)$ is the generalized potential of the Langevin theory we aim to study.    Hermitian conjugate of operator $\Om$ is defined as
\be
(\phi, \Om^\dagger \psi)_{ U} \equiv (\Om \phi, \psi)_{ U} 
= (\psi, \Om \phi )_{ U} .  
\label{O-O-dagger-relation}
\ee
Because we only consider real functions and operators, the inner product is symmetric.  Using integration by parts, we easily see that the Hermitian conjugate of $\partial_i$ is 
\be
\partial_i^\dagger = - \partial_i - (\partial_i U).
\ee
Clearly this also means 
\be
- \partial_i = ( \partial_i + (\partial_i U))^\dagger.
\ee

For many cases, $U(\xv) \rightarrow +\infty$ as $\xv \rightarrow \infty$, so the choice of weight function $e^{U(\xv)}$ of this inner product defined in Eq.~(\ref{inner-prod-def}) is quite unconventional. In order for the norm of a function $(\phi, \phi)$ to be finite, $\phi(\xv)$ needs to decay to zero at least as fast as $e^{-U(\xv)/2}$.  This condition is satisfied by all physically relevant distribution functions which decay as $e^{-U(\xv)}$.  

We decompose ${\mathcal L}_{\rm FP}$  into a part ${\mathcal L}^H$ linear in $B_{ij}$ and another part ${\mathcal L}^A$ linear in $Q_{ij}$:
\begin{subequations}
\label{L_FP-decomposition-0}
\ba
{\mathcal L}_{\rm FP} &=& {\mathcal L}^H + {\mathcal L}^A
= -   \partial_i L_{ij} \partial_j^\dagger,
\label{L_FP-decomposition-1} \\
{\mathcal L}^H &=& -  \partial_i B_{ij} \partial_j^\dagger = {{\mathcal L}^H}^\dagger,\\
{\mathcal L}^A &=& - \partial_i Q_{ij} \partial_j^\dagger  = -{{\mathcal L}^A}^\dagger.   
\ea
\end{subequations}
Hence ${\mathcal L}^H$ is Hermitian whereas ${\mathcal L}^A$ is {\em anti-Hermitian}.  Since $B_{ij}$ is non-negative, the Hermitian operator ${\mathcal L}^H$ is also non-negative, which means that none of its eigenvalues is negative.   

Likewise the  current (\ref{prob-current-2}) can be also decomposed:
\begin{subequations}
\label{J-decomp}
\ba
j_i &=& j_i^H + j_i^A, \\
j_i^H &=& - B_{ij} (\partial_j + (\partial_j U)) p
= B_{ij} \partial_j^\dagger p,
\label{J^I-B}\\
j_i^A &=& Q_{ij} \partial_j^\dagger p + \partial_j (Q_{ij} p).  
 \label{J^I-Q}
\ea
\end{subequations}
 Since $\Bm$ and $\Qm$ do not mix under NTV, these decompositions of ${\mathcal L}_{\rm FP} $ and $j_i$ are covariant.   Note that at the steady state $\partial_i^\dagger p^S = - (\partial_i + \partial_i U) e^{- U}= 0$, and hence $j_i^H $ vanishes, but $j_i^A$ does not need to vanish at stationarity.  

In general the evolution operator $e^{ t {\mathcal L}_{\rm FP}} $ of the Fokker-Planck theory is neither unitary nor Hermitian.  There are however two particularly nice limits of the theory.  If $B_{ij} \rightarrow 0$, then ${\mathcal L}_{\rm FP} \rightarrow {\mathcal L} ^A$, which is anti-Hermitian.  The resulting evolution operator $e^{t {\mathcal L}^A}$ is deterministic and unitary with imaginary eigenvalues, corresponding to oscillatory dynamics.  A simple example of this is Hamiltonian dynamics, where ${\mathcal L}^A$ is just the Liouville operator.  Note, however, this unitary theory is very different from the deterministic irreversible theory shown in Eq.~(\ref{determ-limit-all}).  On the other hand, if $Q_{ij}\rightarrow 0$, then ${\mathcal L}_{\rm FP} \rightarrow {\mathcal L}^H $ and the evolution operator $e^{t {\mathcal L}^H}$ is Hermitian, with real and non-positive engeivatlues. The system relaxes towards steady state without oscillation.  If neither {${\mathcal L}^H$} nor ${\mathcal L}^A$ vanishes, then we may try to treat either {${\mathcal L}^H$} or ${\mathcal L}^A$ as perturbation.  The fact that {${\mathcal L}^H, {\mathcal L}^A$} are respectively Hermitian and anti-Hermitian makes the expansion particularly convenient.  

\vspace{-3mm}
\section{Time-Reversal and Detailed Balance} 
\label{sec:DB}
Detailed balance is a reflection of time-reversal symmetry of the microscopic dynamics.  To discuss time-reversal symmetry, we choose slow variables such that each component has definite time-parity $\varepsilon_i = 1, \,{\rm or}\, -1$. Under time-reversal, we have $x_i \rightarrow \varepsilon_i x_i$. The vector $\xv$ is time-reversed to $\xv^*$ with components $\varepsilon_i x_i$.  For Hamiltonian systems, we have $\xv = (\qv, \pv)$ where $\qv, \pv$ are respectively canonical coordinates and momenta.  Hence we have $\xv^*=(\qv, -\pv)$.   As a consequence the integral measure $d^n \xv$ is also invariant under time-reversal, i.e. $d^n \xv = d^n \xv^*$.  A stationary Markov process is said to be reversible, or satisfy detailed balance, if its steady-state two-time joint {PDF} satisfy~\cite{Gardiner-book,Risken-FP}
\be
p_2^S(\xv_1, t;  \xv_0, 0) =   p_2^S(\xv^*_0, t; \xv^*_1, 0).
 \label{reversible-condition}
\ee  
{(Here we assume that there is no magnetic field or other external field which breaks time-reversal symmetry explicitly, so that the subtle difference between reversibility and detailed balance does not arise.)}  The steady state then becomes the {\em equilibrium state}.  

Suppose the system start from a state $\xv_0$ at $t = 0$, the initial probability {density} is just $p(\xv, 0) = \delta(\xv - \xv_0)$, whereas the probability density $p(\xv_1, t| \xv_0, 0) $ at time $t$, conditioned on the initial position $\xv_0$, is given by 
\ba
p(\xv_1, t| \xv_0, 0) 
&=& e^{ t L_{\rm FP}(\xv_1)} \delta(\xv_1 - \xv_0)
\label{cond-P-integral-1} \\
&=&  \int d^n\xv \, \delta (\xv - \xv_1)
  e^{ t L_{\rm FP}(\xv)} \delta(\xv - \xv_0).
\nonumber 
 \ea
Note that if we integrate $p(\xv_1, t| \xv_0, 0) $ over $\xv_1$, we obtain unity as it should be.  Now because of the Markovian property, the steady-state two-time joint {PDF} $p_2^S(\xv_1, t;  \xv_0, 0)$ can be obtained as
\be
p_2^S(\xv_1, t;  \xv_0, 0) =
 p(\xv_1, t| \xv_0, 0) \,e^{-U(\xv_0)}.
 \label{p_2-p-transition}
\ee

\subsection{Time-reversal}

Starting from Eq.~(\ref{cond-P-integral-1}) we can show:
 \ba
 e^{U(\xv_1)}  p(\xv_1, t  \!\! \!\!  &|&\!\! \!\!   \xv_0, 0) 
  \nonumber\\
&=& 
\int \!\! d^n\xv \,\,  e^{U(\xv)} \delta(\xv - \xv_1)
 e^{t L_{\rm FP}(\xv)} \delta(\xv - \xv_0)
\nonumber\\
 &=& \left(\delta(\xv - \xv_1) , 
e^{t L_{\rm FP}(\xv) } \delta(\xv - \xv_0) \right)_U 
\nonumber\\
&=&\left( \delta(\xv - \xv_0) ,
e^{t L_{\rm FP}^\dagger(\xv) }  \delta(\xv - \xv_1) \right)_{ U} 
 \nonumber\\
  &=& \left( \delta(\xv^* - \xv^*_0), 
  e^{t L_{\rm FP}^\dagger(\xv) } 
  \delta(\xv^* - \xv^*_1)  \right)_{ U} 
 \nonumber\\
&=& \int \!\!  d^n\xv\, e^{U(\xv)} \delta(\xv^* - \xv^*_0) 
  e^{t L_{\rm FP}^\dagger(\xv) } 
  \delta(\xv^* - \xv^*_1)
\nonumber\\
&=& \int \!\!  d^n\xv\, e^{U( \xv^*)} \delta(\xv - \xv^*_0) 
  e^{t L_{\rm FP}^\dagger(\xv^*) } 
  \delta(\xv - \xv^*_1). 
\nonumber\\
 \label{xxx} 
\ea 
In the second and third equalities, we have used respectively definitions (\ref{inner-prod-def}) and (\ref{O-O-dagger-relation}).  In the fourth equality, we have used the time-reversal invariance property of delta function $\delta(\xv) = \delta (\xv^*)$.  In the sixth equality, we have transformed the dummy variable $\xv \rightarrow \xv^*$.  This does not change the integral, since the volume measure $d^n\xv$ is invariant under time-reversal.  Note that the weight function is also transformed from $U(\xv)$ to $ U(\xv^*)$.   

We  define  a new Fokker-Planck operator $\tilde{L}_{\rm FP}(\xv)$ as\xing{}
\ba
\tilde{L}_{\rm FP}(\xv) &\equiv& L_{\rm FP}^\dagger(\xv^*) 
= \partial_i^* L_{ji}(\xv^*) 
(\partial_j^* + (\partial_j^* U(\xv^* ) ) )
\nonumber\\
 &=& \frac{\partial}{\partial x^*_i} L_{ji}(\xv^*) 
\left( \frac{\partial}{\partial x^*_j} 
+ \left( \frac{\partial}{\partial x^*_j} U(\xv^* ) \right) \right),
\nonumber\\
 \label{L-tilde-L} 
 \label{L_FP-tilde-def}
\ea  
where $ x^*_i = \varepsilon_i x_i$, and $\partial^*_i 
= \varepsilon_i \partial/\partial  x_i $.  Further defining two functions $\tilde L_{ij}(\xv)$ and  $\tilde U(\xv)$ via
\ba
\tilde L_{ij}(\xv) \equiv  \varepsilon_i
 L_{ji}(\xv^*) \varepsilon_j, \quad
\tilde U(\xv) \equiv U(\xv^*), 
\label{A-tilde-U-tilde-def}
\ea
we can write $\tilde{L}_{\rm FP}(\xv) $ in the following form:
\ba
\tilde{L}_{\rm FP}(\xv) = \partial_i \tilde L_{ij} (\partial_j + \partial_j \tilde U(\xv)). 
\label{L_FP-tilde-def}
\ea
which has the standard form of Fokker-Planck operator (c.f.  Eq.~(\ref{L_FP-def})), but with $\Lm, U$ replaced by $\tilde \Lm$ and $\tilde U$.  Recall the decomposition (\ref{L-B-Q-decomp}), Eq.~(\ref{A-tilde-U-tilde-def}) also implies
\be
\tilde B_{ij} (\xv) =  \varepsilon_i B_{ij}(\xv^*) \varepsilon_j, \quad
\tilde Q_{ij} (\xv) = - \varepsilon_i Q_{ij}(\xv^*) \varepsilon_j. 
\ee


Let us study the steady state of the Langevin process defined by  $\tilde{L}_{\rm FP}(\xv)$, which for now shall be referred as the {\em tilde process}.  The steady state distribution is 
\be
\tilde p^S(\xv) = e^{-\tilde U( \xv)} = e^{-\tilde U( \xv^*)} = p^S(\xv^*),
\label{P^S-relation}
\ee 
which is the time-reversal of the steady state of the original process $e^{-U(\xv)}$.   The steady state probability current of the tilde process can be obtained from Eq.~(\ref{F^S-2}):
\be
\tilde j^S_i(\xv) = \partial_j (\tilde Q_{ij}(\xv) e^{-\tilde U(\xv)} ),
\label{tilde-J^S}
\ee
which is related to the original steady state current via  
\be
\tilde j^S_i(\xv^*) = - \varepsilon_i j^S_i(\xv).
\label{J^S-relation}
\ee  

Equation (\ref{xxx}) can now be rewritten as 
\ba
e^{U(\xv_1)}  p(\xv_1, t |  \xv_0, 0) 
\!\!  &=& \!\!\!\! \int_{\xv} \!\! e^{\tilde U(\xv)} \delta (\xv - \xv^*_0) 
e^{t \tilde L_{\rm FP} (\xv)} \delta (\xv - \xv^*_1)
\nonumber\\
&=& e^{\tilde U (\xv^*_0)} \tilde p(\xv^*_0, t | \xv^*_1, 0),
\ea
where $\tilde p(\,\cdot \, | \,\cdot\,)$ is the transition probability {density} of the tilde process.  Further multiplying both sides by $e^{ - \tilde U(\xv_1^*) - U (\xv_0)}$, we obtain\xing{}
\ba
 p(\xv_1, t | \xv_0, 0) e^{- U(\xv_0)}  &=& 
 \tilde p(\xv^*_0, t | \xv^*_1, 0) e^{- \tilde U (\xv^*_1)}. 
 \quad\quad
\ea 
According to Eq.~(\ref{p_2-p-transition}), this can be further written as an equality between the steady state two-time joint pdf of the original process and the tilde process:
\ba
p^S_2(\xv_1, t;  \xv_0, 0) &=&   \tilde p^S_2(\xv^*_0, t; \xv^*_1, 0).  
\label{reversibility-Markov-cond}
\ea
Eqs.~(\ref{P^S-relation}), (\ref{J^S-relation}), and (\ref{reversibility-Markov-cond}) demonstrate the physical significance of the tilde process defined by $\tilde L_{\rm FP}$: It is the {\em macroscopic time-reversal} of original process corresponding to $L_{\rm FP}$, because all macroscopic properties are reversed.  

\subsection{Covariant Formulation of Detailed Balance}
Recall that detailed balance is defined by the condition (\ref{reversible-condition}).  Combining this with Eq.~(\ref{reversibility-Markov-cond}) we find  
\ba
&& p^S_2(\xv_1, t;  \xv_0, 0) =    \tilde p^S_2(\xv_1, t;  \xv_0, 0),
\nonumber
\ea
which in turn implies 
\ba
&& \tilde{\mathcal L}_{\rm FP}(\xv) 
  = {\mathcal L}_{\rm FP}(\xv).
  \label{L-L-dagger-relation}
\ea
In view of Eqs.~(\ref{L_FP-tilde-def}), this is further equivalent to\xing{}
\begin{subequations}
\label{reversibility-0}
\ba
\tilde U(\xv) &\equiv& U(\xv^*) =  U(\xv),
\label{reversibility-U-0} \\
\tilde  L_{ij} (\xv) &\equiv& \varepsilon_i L_{ji}(\xv^*) \varepsilon_j 
=  L_{ij} (\xv).  
\label{reversibility-L-0}
\ea
Equations~(\ref{reversibility-L-0}) are further equivalent to 
\ba
\tilde  B_{ij} (\xv) &\equiv& \varepsilon_i B_{ij}(\xv^*)  
 \varepsilon_j =  B_{ij}(\xv),
\label{reversibility-B-0}
\\
\tilde Q_{ij} (\xv) &\equiv& - \varepsilon_i Q_{ij}(\xv^*)  
\varepsilon_j =  Q_{ij}(\xv). 
\label{reversibility-Q-0}
\ea
\end{subequations}
 Equations (\ref{reversibility-0}) are the necessary and sufficient conditions for {\em detailed balance} of Langevin dynamics (\ref{NL-Langevin-standard}). In the setting of linear response theory, these conditions are better known as Onsager-Casimir reciprocal symmetry of the kinetic coefficients $L_{ij}$~\cite{Onsager-reciprocal-1931,Casimir-reciprocal-1945}.  They agree with Eq.~(2.72) of reference \cite{Eyink-1996}, and also with those derived in Gardiner Sec. 5.3.5~\cite{Gardiner-book}.  Note however Gardiner's (5.3.53 iii) is much more complicated and less transparent.  
 
 Combining Eqs.~(\ref{J^S-relation}) and (\ref{reversibility-Q-0}), we find that for reversible Langevin dynamics, the stationary probability current transforms  under time-reversal as\xing{}
\be
  j^S_i(\xv^*) = - \epsilon_i \,j^S_i(\xv) . 
\ee


We can now explicitly show that the conditions of DB (\ref{reversibility-0}) are covariant under NTVs that respect time-reversal symmetry.  But the latter, we mean that the each $y_a$ of the new variables $\yv = (y_1, \ldots, y_n)$ also has definite time-parity $\varepsilon_a$, and further satisfy $y_a^*(\xv^*) = y_a(\xv)$.   But this also means $\partial y_a(\xv)/\partial x_i = \varepsilon_a \partial y_a(\xv^*)/\partial x_i \, \varepsilon_i$, and hence the Jacobian is invariant under time reversal: $J(\xv^*) = J(\xv)$. Using these together with Eqs.~(\ref{covariant-rules-U}) and (\ref{covariant-rules-3}), we can easily see that Eqs.~(\ref{reversibility-U-0}) and (\ref{reversibility-L-0}) reduce to 
\begin{subequations}
\ba
\tilde U'(\yv) &\equiv& U'(\yv^*) =  U'(\yv),
\label{reversibility-U-0-1} \\
\tilde  L'_{ab} (\yv) &\equiv& \varepsilon_a L_{ba}(\yv^*)\varepsilon_b,
\label{reversibility-L-0-1} 
\ea
\end{subequations}
which are just the conditions of DB in the new variables. Hence DB keeps its identity during NTV, as we expected.


\subsection{Monotonic decrease of free energy }
We can show that a functional of pdf $p(\xv, t)$ monotonically decreases as a function of time.  For reversible Markov process which satisfies DB, this result can be understood as a reflection of the second law of thermodynamics, which dictates that the total entropy of an isolated system can only increase. We will call this functional {\em free energy}, which is defined as 
\ba
F[p(t)] &\equiv& T\, \int d\xv\, p(\xv,t) \left[
U(\xv) + \log p(\xv, t)  \right], \quad
\ea
where $T$ (temperature) is just a constant of proportionality and plays no significant role here.   The time derivative of $F[p(t)] $ can be calculated using the Fokker-Planck equation:
\ba
\frac{1}{T} \frac{dF}{dt} &=& \int d\xv 
\left[ U(\xv) +  \log p(\xv, t)  \right] (\partial_t p)
\nonumber\\
&=& -  \int d\xv \, \left[ U(\xv) + \log p(\xv, t)  \right]
\partial_i L_{ij}\partial_j^\dagger p
\nonumber\\
&=&   \int d\xv \, \{ \partial_i  \left[ U(\xv) + \log p(\xv, t)  \right] \}
L_{ij}\partial_j^\dagger p
\nonumber\\
&=&   \int d\xv \, p^{-1} ((\partial_i U) + \partial_i) p) 
L_{ij}\partial_j^\dagger p
\nonumber\\
&=& -  \int d\xv \,p^{-1} 
(\partial_i^\dagger p) L_{ij} (\partial_j^\dagger p)
\nonumber\\
&=& -  \int d\xv \,p^{-1} 
(\partial_i^\dagger p) B_{ij} (\partial_j^\dagger p)
\leq 0, 
\label{rate-F}
\ea
where in the last line, we have used the facts that the antisymmetric part of $L_{ij}$ does not contribute to the quadratic form, and that the matrix $B_{ij}$ is semi-positive definite.  Hence the free energy decreases over time.  

If $\Bm$ is positive definite, the above inequality is sufficient to guarantee that the system converges to a unique equilibrium state with minimal free energy.  Furthermore, using Eq.~(\ref{J-decomp}), we can also express Eq.~(\ref{rate-F}) in terms of the Hermitian probability current $j^H$:
\ba
\frac{1}{T} \frac{dF}{dt} &=& - \int d\xv \,
P^{-1} j^H_i B^{-1}_{ij} j^H_j. 
\ea
Hence we find that the Hermitian current $j^H$ but not the anti-Hermitian current $j^A$ contributes to entropy production (c.f. Eqs.~(\ref{J-decomp})).   It is then appropriate to call $\Qm$ the reactive couplings whereas $\Bm$ dissipative couplings. This is consistent with Eqs.~(\ref{reversibility-Q-0}) and (\ref{reversibility-B-0}) which say that $\Qm$ couple even variables to odd variables, while $\Bm$ couple variables with the same signature of time-reversal.   For systems without DB, however, both ${\mathcal L}^H$ and ${\mathcal L}^A$ may contribute to dissipation in general and the term {\em reactive} or {\em conservative} can not be applied to $\Qm$.


\section{Applications}
In this section, we discuss two simple applications of our theory.  We will start from the simple and well-known case of linear response theory, and then discuss a slightly more complicated case of weakly damped classical Hamiltonian system.  Both systems however are reversible and have additive noises.  More complicated cases (without detailed balance and with multiplicative noises) will be discussed in future publications.  

\subsection{Linear Response Theory}
The simplest case is that all kinetic coefficients $L_{ij}$ are constants, and the generalized potential is quadratic: 
\begin{subequations}
\be
U(\xv) = \frac{1}{2} s_{ij} x_i x_j. 
\ee 
The Langevin equations then become linear:
\ba
&&\dot x_i + ( B_{ij} + Q_{ij} )s_{jk} x_k = \eta_{i}, \\
&& \langle \eta_i (t_1) \eta_j(t_2)\rangle = 2 B_{ij} \delta(t_1 - t_2),
\ea
where the noises are related to $\xi_{\alpha}$ defined in Eq.~(\ref{NL-Langevin-1-1}) via
\be
\eta_j(t) = b_{j \alpha} \xi_{\alpha}(t). 
\ee
\end{subequations}
The detailed balance conditions (\ref{reversibility-0}) guarantee that  $B_{ij}$ couple only variables with same time-reversal symmetry, and  $Q_{ij}$ only couple variables with opposite time-reversal symmetry.  These are precisely the reciprocal symmetry discovered by Onsager~\cite{Onsager-reciprocal-1931} and Casimir~\cite{Casimir-reciprocal-1945} long ago.  

\subsection{Weakly Damped Hamiltonian System}
The slightly more complex case is that $L_{ij}$ remain constants, but $U(\xv)$ becomes an arbitrary function bound from below.  One of the simplest realization of this case is a one-dimensional  classical Harmonic system damped by weak ambient noises.   The slow variables are $\xv = (q, p)$, where $q,p$ are the canonical coordinate and momentum.  Let $H(q,p) = p^2/2m + V(q)$ be the Hamiltonian.  The steady state is the equilibrium state with Gibbs-Boltzmann distribution $p^S(q,p) = e^{-\beta H(q,p) + \beta F(T)}$, where $F(T)$ is the Free energy.  Hence the generalized potential is 
\be
U = \beta H - \beta F
= \beta \left(  p^2/2m + V(q) - F \right) . 
\label{U-H-form}
\ee  
The Langevin equations are 
\begin{subequations}
\label{Langevin-Hamiltonian}
\ba
\dot q &=& \frac{\partial H}{\partial p} =  \frac{p} {m} , \\
\dot p &=& - {\gamma} \frac{\partial H}{\partial p}
- \frac{\partial H }{\partial q} + \eta (t) 
\nonumber\\
&=& - \gamma \dot q - \partial_{q} V(\qv) + \eta (t).
\ea
\end{subequations}
These equations can be rewritten as the standard form:
\ba
 \begin{pmatrix}
\dot q  \\ \dot p
\end{pmatrix}
 +  \begin{pmatrix}
0 & -T \\
T & T \gamma
\end{pmatrix} 
\begin{pmatrix}
\partial_q U \\
\partial_p U 
\end{pmatrix}
= \begin{pmatrix}
0 \\ \eta (t)
\end{pmatrix}.
\ea
From this we read off the matrix of kinetic coefficients:
\be
\Lm =  \begin{pmatrix}
0 & -T \\
T & T \gamma
\end{pmatrix},  \,\,
\Bm =  \begin{pmatrix}
0 & 0  \\
0  & T \gamma
\end{pmatrix}, \,\,
\Qm =  \begin{pmatrix}
0 & -T \\
T & 0
\end{pmatrix}. 
\ee 
The detailed balance conditions (\ref{reversibility-0}) can be easily verified.
 The noise variance is given by the Einstein relation
\be 
\langle \eta (t_1) \eta (t_2)\rangle 
= 2 T \gamma \, \delta(t_1 - t_2). 
\ee
Note that $L_{ij}$ are proportional, whereas $U$ is anti-proportional, to temperature $T$, as we have claimed in Sec.~\ref{sec:determinstic}.  As a consequence, Eqs.~(\ref{Langevin-Hamiltonian}) depend on temperature only through the noise variance (assuming that the friction coefficient $\gamma$ is independent of $T$).  If we take the zero temperature limit, we obtain a set of deterministic irreversible equations:
\begin{subequations}
\ba
\dot q &=&   \frac{p} {m} , \\
\dot p &=& - \gamma \dot q - \partial_{q} V(\qv).
\ea
\end{subequations}
This is the deterministic limit we discussed in Sec.~\ref{sec:determinstic}.   
As discussed there, the same set of equations can also be obtained from Eqs.~(\ref{Langevin-Hamiltonian}) in the thermodynamic limit, where the particle is very massive, so that the noise can be neglected.  

As discussed in Sec.~\ref{sec:unitary-Hermitian}, if we let the symmetric part of kinetic coefficients go to zero, $\gamma \rightarrow 0$, we obtain a unitary and deterministic dynamics.  The Langevin equations then become
\begin{subequations}
\ba
\dot q &=&   \frac{p} {m} , \\
\dot p &=& - \partial_{q} V(\qv).
\ea
\end{subequations}
But these are just the Hamiltonian dynamics without damping.  Hence the unitary limit corresponds to the limit of vanishing friction and noise.



\vspace{5mm}
\section{Conclusion}
\label{sec:conclusion}
\vspace{-3mm}
 We conclude our work with a few comments on the choices of stochastic calculus.   It is known~\cite{Gardiner-book,Jacobs-book} that Ito-Langevin equation (\ref{NL-Langevin-1-1}) is mathematically equivalent to  Stratonovich-Langevin equation:
\ba
\dot x_i = F^S_i(\xv,t) +  b_{i \alpha}(\xv) \circ \xi_\alpha(t),
\label{NL-Langevin-Strato} 
\ea 
if we impose the following relation between $F^S_i\ $ and $F_i$:
\ba
F^S_i = F_i - \frac{1}{2} b_{j \alpha} \partial _j b_{i \alpha}.
\label{NF^S-F} 
\ea
Here in Eq.~(\ref{NL-Langevin-Strato}) the product $b_{i \alpha}(\xv) \circ \xi_\alpha(t)$ is interpreted in the sense of Stratonovich.   Hence our covariant Langevin equation (\ref{NL-Langevin-standard}) can also be represented as a {Stratonovich}-Langevin equation.  One can further show that $F^S_i(\xv,t)$ in Eq.~(\ref{NL-Langevin-Strato}) transforms as contra-variant vector under NTV.  In fact this simple transformation law for $F^S_i(\xv,t)$ has been deemed as a major advantage of Stratonovich over Ito. Of course, in this work we have demonstrated that Ito-Langevin is also fully covariant as long as it is parameterized in terms of $U, \Lm$.  So Ito-Langevin is at least as convenient as Stratonovich-Langevin.  

 Stratonovich-Langevin however has inconvenient features.  Its drift term $F^S_i(\xv,t)$, though covariant, is connected to observables $U(\xv)$ and $\Lm$ in a more complex way, and hence conditions of DB become obscure.   More importantly, as pointed out by Graham~\cite{Graham-covariance-Langevin-Ito}, $F^S_i(\xv,t)$ is changed by $\xv$-dependent right-rotation of noises $b_{i\alpha}(\xv)$, and hence has lower symmetry than Ito-Langevin.  Qualitatively, it indicates that the Stratonovich-Langevin equation is sensitive to details of fast variables, a rather strange feature.  Finally, as is well known, the Stratonovich-Langevin equation (\ref{NL-Langevin-Strato}) ``looks into the future'', which makes numerical studies very inconvenient and violates the principle of causality.  Quite obviously, all these comments apply to other non-Ito schemes as well.  Our conclusion is that the formalism developed here has higher symmetry and clearer physical meanings, as well as much simpler behaviors under general NTV, and hence is a more natural formalism for non-linear Langevin dynamics with multiplicative white noises.   

 The authors acknowledge support from NSFC via grant 11674217(X.X.) and 11675017(Z.C.T.), as weak as Shanghai Municipal Science and Technology Major Project (Grant No.2019SHZDZX01). 
 X.X. also thanks additional support from a Shanghai Talent Program.

\pagebreak
\begin{widetext}

\appendix
\section{Covariance of Langevin  and Fokker-Planck Equations}
\label{sec:app_A}
Our goal is to prove that under the rules Eqs.~(\ref{transform-U-L-b}), the non-linear Langevin equation,  the Fokker-Planck operator, probability current, and FPE transform respectively as Eqs.~(\ref{dy-Ito-1-2}) and (\ref{transformed-FP}).   For this purpose, we first need to prove a few useful identities.  


\subsection{Jacobi's formula}
Let us first establish two very useful  identities about Jacobian.  Let $\Am$ be a nonsingular square matrix, with determinant $\det \Am$, inverse $\Am^{-1}$, then {\em Jacobi's formula} says  (here $d$ denotes differential)
\be
d \log \det \Am = \Tr\, \Am^{-1} d \Am. 
\ee 
Applying this formula to matrix $\partial y_a/\partial x_i$ which has determinant $J$ and inverse $\partial x_i /\partial y_a$, we have
\ba
 d \log J &=&  J^{-1} d J
=  \left( \frac{\partial x_i}{ \partial y_a}\right)
 d \left(\frac { \partial y_a} {\partial x_i}\right). 
 \ea
Hence $\forall \, j$ we have
\ba
 J^{-1}\frac{\partial   J}{\partial x_j}  
&=& \left( \frac{\partial x_i}{ \partial y_a}\right)
 \frac{\partial}{\partial x_j} \left(\frac { \partial y_a} {\partial x_i}\right)
\nonumber\\
&=& \left(  \frac{\partial x_i}{ \partial y_a}\right)
 \left(\frac { \partial^2 y_a} {\partial x_j\partial x_i}\right)
 = \frac{\partial}{\partial y_a} \frac { \partial y_a} {\partial x_j},
\label{Jacobi-equality}
\ea
from which we further prove a useful identity:
\be
\frac{\partial}{\partial y_a} \left( 
\frac{\partial y_a}{\partial x_i}  J^{-1}  \right) = 0.  
\label{identity-useful-1}
\ee
Swapping the roles of $\xv$ and $\yv$, we obtain a result that is reciprocal to Eq.~(\ref{identity-useful-1}):
\be
\frac{\partial}{\partial x_i} \left( 
\frac{\partial x_i}{\partial y_a}  J  \right) = 0.  
\label{identity-useful-1-1}
\ee
Note that in the above when taking partial derivative with respect to $x_i$ ($y_a$) it is always understood that all other $x_j, \,\, j \neq i$ ($y_b, \,\, b \neq a$) are fixed.  Eqs.~(\ref{identity-useful-1}) and (\ref{identity-useful-1-1}) will be very useful below.

\subsection{Proof of Eq.~(\ref{dy-Ito-1-2})}

Let us prove Eq.~(\ref{dy-Ito-1-2}).  Let us rewrite Eq.~(\ref{NL-Langevin-standard}) and Eq.~(\ref{dy-Ito-1-2}) in an equivalent form which is preferred by mathematician:
\begin{subequations}
\begin{align}
d x_i &= ( - L_{ij} \partial_j U + \partial_j L_{ij} ) dt 
+ b_{i\alpha}    d W_{\alpha}(t),
\tag{2.12'}\\[8pt]
d y_a &=  \left(  -  L'_{ab} \partial_b U' 
+ \partial_b L'_{ab}  \right) dt 
+ b'_{a \alpha}  d W_\alpha(t). 
\tag{2.19'}
\end{align}
\end{subequations}
Here $dW_{\alpha}(t)$ are differential of Wiener's processes, and obey Ito's rule~\cite{Jacobs-book}:
\be
 d W_{\alpha}(t) dW_{\beta}(t)
= \delta_{\alpha \beta} dt .
\label{Ito-rule}
\ee
Let $y(\xv)$ be a function of $\xv$, Ito's formula~\cite{Gardiner-book,Jacobs-book} relates the differential of $\yv$ to that of $\xv$:
\be
d y = \frac{\partial y}{\partial x_i} d x_i 
+ \frac{1}{2} \frac{\partial^2 y}
{\partial x_i \partial x_j} d x_i dx_j. 
\label{Ito-formula}
\ee

We will derive Eq.~(2.19') from Eq.~(2.12') using rules Eqs.~(\ref{transform-U-L-b}) and  {\em Ito's formula} Eq.~(\ref{Ito-formula}).  We use Eq.~(2.12') to rewrite $dx_i$ and $d x_j$ in Eq.~(\ref{Ito-formula}) in terms of $dt$ and $dW(t)$.  For the quadratic term $dx_i dx_j$, however, we only need to keep terms proportional to $(d W)^2 \sim dt$. Further using Ito's rule (\ref{Ito-rule}), we obtain:
\ba
d y_a &=& -  \frac{\partial y_a}{\partial x_i} L_{ij} (\partial_j U) dt 
 + \frac{\partial y_a}{\partial x_i}  \partial_j L_{ij} dt
+ \frac{\partial^2 y_a}{\partial x_i \partial x_j} B_{ij}dt
+  \frac{\partial y_a}{\partial x_i} b_{i \alpha} d W_\alpha. 
\label{dy_a-1}
\ea
Using Eqs.~(\ref{covariant-rules-U}) and (\ref{covariant-rules-3}) as well as the chain rule, the first term in the RHS of Eq.~(\ref{dy_a-1}) can be rewritten as $ -  L'_{ab} (\partial_b  U' - \partial_b  \log j) d t$.  Using Eq.~(\ref{covariant-rules-b}), the last term can be rewritten as $b'_{a \alpha} d W_\alpha$.  Hence we have
\ba
d y_a &=&  \left(  -  L'_{ab} \partial_b U' 
+ \partial_b L'_{ab}  \right) dt 
+ b'_{a \alpha} d W_\alpha
+ \Psi \, dt, 
\label{dx-Ito-2}\\
\Psi & \equiv &  L'_{ab}\partial_b \log j
+ \frac{\partial y_a}{\partial x_i}  \partial_j L_{ij}
- \partial_b L'_{ab}
+ \frac{\partial^2 y_a}{\partial x_i \partial x_j} B_{ij}.
\label{Psi-def}
\ea
Note that Eq.~(\ref{dx-Ito-2}) differs from {Eq.~(2.19') } only by the term $\Psi dt$, which will be shown to vanish identically.  The first term in $\Psi$ can be calculated using Eq.~(\ref{Jacobi-equality}) and chain rule as well as commutativity of derivatives $\partial_i, \partial_j$:
\ba
  L'_{ab}\partial_b \log j 
 &=& L'_{ab} \frac{\partial x_i}{\partial y_c} 
  \frac{\partial }{\partial y_b} \frac{\partial y_c}{\partial x_i}
= \frac{\partial y_a}{\partial x_j}
L_{jk} \frac{\partial y_b}{\partial x_k}
\frac{\partial x_i}{\partial y_c}
  \frac{\partial }{\partial y_b} \frac{\partial y_c}{\partial x_i}
\nonumber\\
&=& \frac{\partial y_a}{\partial x_j}
L_{jk}\frac{\partial x_i}{\partial y_c}
 \frac{\partial }{\partial x_k} \frac{\partial y_c}{\partial x_i}
= \frac{\partial y_a}{\partial x_j}
L_{jk}\frac{\partial x_i}{\partial y_c}
 \frac{\partial }{\partial x_i} \frac{\partial y_c}{\partial x_k}
\nonumber\\
&=& \frac{\partial y_a}{\partial x_j}
L_{jk}
 \frac{\partial }{\partial y_c} \frac{\partial y_c}{\partial x_k}
= \frac{\partial y_a}{\partial x_i}
L_{ij}
 \frac{\partial }{\partial y_b} \frac{\partial y_b}{\partial x_j}.
\label{intermediate-1}
   \ea
Using Eq.~(\ref{covariant-rules-3}), negative the third term in $\Psi$ can be rewritten as:
\ba
\partial_b L'_{ab} &=& 
\frac{ \partial}{\partial y_b} \left( 
\frac{\partial y_a}{\partial x_i} 
L_{ij} \frac{\partial y_b}{\partial x_j}
\right)
\nonumber\\
&=& \frac{\partial y_a}{\partial x_i} 
 \frac{\partial y_b}{\partial x_j} \frac{ \partial}{\partial y_b}  L_{ij}
 + \frac{ \partial}{\partial y_b}  \frac{\partial y_a}{\partial x_i} 
 L_{ij} \frac{\partial y_b}{\partial x_j} 
 +   \frac{\partial y_a}{\partial x_i} 
L_{ij}  \frac{ \partial}{\partial y_b} 
 \frac{\partial y_b}{\partial x_j} 
\nonumber\\
&=&  \frac{\partial y_a}{\partial x_i} 
{\partial_j}  L_{ij}
+ \left(  \frac{\partial^2 y_a}{\partial x_j \partial x_i} \right) L_{ij}
+   \frac{\partial y_a}{\partial x_i} 
L_{ij}  \frac{ \partial}{\partial y_b} 
 \frac{\partial y_b}{\partial x_j} . 
\nonumber\\
&=&  \frac{\partial y_a}{\partial x_i }  \partial_j L_{ij}
+ \left(  \frac{\partial^2 y_a}{\partial x_j \partial x_i} \right) B_{ij}
+   \frac{\partial y_a}{\partial x_i} 
L_{ij}  \frac{ \partial}{\partial y_b} 
 \frac{\partial y_b}{\partial x_j} . 
\label{intermediate-2}
 \ea
In the final step, we have used the fact that  $\frac{\partial^2 y_a}{\partial x_i \partial x_j} $ is symmetric in $i,j$ and hence $\frac{\partial^2 y_a}{\partial x_i \partial x_j}  Q_{ij} = 0$. 

Substituting Eqs.~(\ref{intermediate-1}) and (\ref{intermediate-2}) back into Eq.~(\ref{Psi-def}), we finally see that all terms cancel exactly in RHS, and hence  $\Psi$ vanishes identically.  This means Eq.~(\ref{dx-Ito-2}) reduces to Eq.~(2.19'), as we expected.

\subsection{Proof of Eq.~(\ref{covariant-rules-5}) }

First, using Eq.~(\ref{identity-useful-1-1}), we immediately obtain the {\em operator identity}:
\be
\frac{\partial}{\partial y_a} 
= J^{-1} 
\frac{\partial}{\partial x_l } \frac{\partial x_l}
{\partial y_a}J 
\label{identity-useful-2-0}
\ee
Also using the chain rule,  (\ref{identity-useful-1-1}) as well as Eq.~(\ref{covariant-rules-U}), we have:
\ba
 \left( \frac{\partial}{\partial y_b} 
+ \frac{\partial U'}{\partial y_b}  \right)
&=& \frac{\partial x_k}{\partial y_b} 
\left(   \frac{\partial}{\partial x_k} 
+ \frac{\partial U'}{\partial x_k}  \right)
 \nonumber\\
&=&  \frac{\partial x_k}{\partial y_b} 
  \left(  \frac{\partial}{\partial x_k}
  + \frac{\partial  U}{\partial x_k}
  + J^{-1} 
  \frac{\partial  j}{\partial x_k}
 \right) 
 \nonumber\\
&=& J^{-1}   \frac{\partial x_k}{\partial y_b} 
  \left( \frac{\partial}{\partial x_k}
  + \frac{\partial  U}{\partial x_k} \right)
  J. 
\label{identity-useful-2}
\ea
Now take the product of Eq.~(\ref{identity-useful-2-0}), Eq.~(\ref{covariant-rules-3}), and  Eq.~(\ref{identity-useful-2}) consecutively.  On the LHS we obtain ${\mathcal L}'_{\rm FP}$ according to Eq.~(\ref{Fokker-Planck-2-c}).  On the RHS we find
\ba
J^{-1}   \frac{\partial}{\partial x_i}
L_{ij} \left( \frac{\partial}{\partial x_i}
 + \frac{\partial U} {\partial x_i}\right)
   J = J^{-1}  
   {\mathcal L}_{\rm FP} J. 
  \label{RHS-prod} 
\ea
Hence we obtain Eq.~(\ref{covariant-rules-5}) as an operator identity.  

  
\subsection{Proof of Eqs.~(\ref{covariant-rules-4}) and (\ref{covariant-rules-4-1})}
Taking the product Eq.~(\ref{covariant-rules-3}) $\times$ Eq.~(\ref{identity-useful-2}) and acting on Eq.~(\ref{covariant-rules-1}), we obtain
\be
- L'_{ab}(\partial'_a + (\partial'_b U')) p' =
-  J^{-1}
\frac{\partial y_a}{\partial x_i} L_{ij}
(\partial_j + (\partial_j U)) p. 
\label{L'-L-1}
\ee
Using Eqs.~(\ref{Q'-Q}) and (\ref{covariant-rules-1}), we can also show 
\ba
 \partial'_b(Q'_{ab} p' )&=& \frac{\partial}{\partial y_b}
\left( \frac{\partial y_a}{\partial x_i} Q_{ij} 
\frac{\partial y_b}{\partial x_j} J^{-1} \, p
\right)
\nonumber\\
&=&  \frac{\partial y_a}{\partial x_i} \frac{\partial y_b}{\partial x_j} J^{-1} 
\frac{\partial}{\partial y_b} Q_{ij} p
\nonumber\\
&+&  \frac{\partial y_a}{\partial x_i} Q_{ij} p
\frac{\partial}{\partial y_b} \frac{\partial y_b}{\partial x_j} J^{-1}
\nonumber\\
&+& Q_{ij}p \frac{\partial y_b}{\partial x_j} J^{-1}
\frac{\partial}{\partial y_b} \frac{\partial y_a}{\partial x_i}.
\ea
In the RHS, the second term vanishes because of Eq.~(\ref{identity-useful-1}).  The third term vanishes because it can be rewritten as 
$Q_{ij} p J^{-1}  \frac{\partial^2 y_a}{\partial x_i\partial x_j}$,
which again vanishes because of the antisymmetry of $Q_{ij}$.  Hence we find that 
\be
 \partial'_b(Q'_{ab} p' ) = J^{-1}
\frac{\partial y_a}{\partial x_i} L_{ij}
\partial_j(Q_{ij} p ). 
\label{L'-L-2}
\ee
Adding up Eqs.~(\ref{L'-L-1}) and (\ref{L'-L-2}), and using Eq.~(\ref{prob-current-2}) and Eq.~(\ref{J'-L'-P'}) , we obtain Eq.~(\ref{covariant-rules-4}):
\begin{align}
j'_a  = J^{-1} (\partial y_a/\partial x_i) j_i. 
\tag{2.16b}
\end{align}
Taking the partial derivative $\partial'_a$ of both sides of Eq.~(\ref{covariant-rules-4}), and using Eq.~(\ref{identity-useful-1}), we easily find Eq.~(\ref{covariant-rules-4-1}):
\begin{align}
\partial'_a j'_a =  J^{-1} \partial_i j_i. 
\tag{2.16c}
\end{align}
At steady state, both sides vanish.  Hence steady state is transformed into a steady state. Because of this, Eq.~(\ref{L'-L-2}) can also be written as
\be
j_a^{'S}
= J^{-1}   \frac{\partial y_a}{\partial x_i} j^S_i. 
\ee






\end{widetext}

\end{document}